\documentclass[sigconf]{acmart}%[15pt]

\usepackage[english]{babel}
\usepackage{blindtext}

\usepackage{graphicx,subfig}
\usepackage{caption}
\usepackage[normalem]{ulem}
\usepackage{algorithm}
\usepackage{algorithmic}
\usepackage{multirow}
\usepackage{amsopn}
\usepackage{etoolbox}
\usepackage{xcolor}
\usepackage{url}
\usepackage{hyperref}
\usepackage{xspace}
\usepackage{indentfirst}
\usepackage{breakurl}
\usepackage{microtype}
\usepackage{color}

\newcommand{\cosar}{{\sc CoIC}\xspace}
\newcommand{\name}{{\sc CoIC}\xspace}

\newcommand{\eg}{\emph{e.g.,}\xspace}

\setcopyright{acmcopyright}
\acmDOI{}

\acmISBN{}

\acmPrice{}

\begin{document}
\title{Immersion on the Edge: A Cooperative Framework \\ for Mobile Immersive Computing}

%\titlenote{Produces the permission block, and copyright information}
%\subtitle{Extended Abstract}

%\author{Paper \# 2, 2 pages}

%\affiliation{
%  \institution{Tsinghua University}
%}
%\email{laizq13@mails.tsinghua.edu.cn}
%2

%Zeqi Lai, Y. Charlie Hu$^*$, Yong Cui, Linhui Sun,Ningwei Dai\\
%\large{Tsinghua University, $^*$Purdue University}

%\numberofauthors{1}
%\author{
%%Paper \#8, 12 Pages
%}

%1
%\author{Zeqi Lai, Yong Cui\footnotemark[1], Ziyi Wang, Xiaoyu Hu}
%1
\author{Zeqi Lai}
\affiliation{
\institution{Tsinghua University}
}
\email{laizq13@mails.tsinghua.edu.cn}
%2
\author{Yong Cui}
\authornote{Corresponding author. This project is supported by National Key R\&D Program of China under Grant 2017YFB1010002, National 863 project (no. 2015AA015701).}
\affiliation{
\institution{Tsinghua University}
}
\email{cuiyong@tsinghua.edu.cn}
%3
\author{Ziyi Wang}
\affiliation{
\institution{Tsinghua University}
}
\email{wangziyi0821@gmail.com}
%4
\author{Xiaoyu Hu}
\affiliation{
\institution{Tsinghua University}
}
\email{chaoese@gmail.com}

%\email{laizq13@mails.tsinghua.edu.cn, cuiyong@tsinghua.edu.cn, {wangziyi0821, chaoese}@gmail.com}

%\saythanks
%\authornote{Yong is the corresponding author.}
%%3
%\author{Ziyi Wang}
%\author{Xiaoyu Hu}
%\affiliation{
%  \institution{Tsinghua University}
%}
%\email{cuiyong@tsinghua.edu.cn}

%\affiliation{
%  \institution{Tsinghua University}
%}
%\email{wangziyi_0821@foxmail.com}
%4

%\affiliation{
%  \institution{Tsinghua University}
%}
%\email{chaoese@gmail.com}

% \author{Firstname Lastname}
% \authornote{Note}
% \affiliation{%
%   \institution{Affiliation}
%   \streetaddress{Address}
%   \city{City}
%   \state{State}
%   \postcode{Zipcode}
% }
% \email{email@domain.com}


%\renewcommand{\shortauthors}{X.et al.}

%\input{abstract}

\copyrightyear{2018}
\acmYear{2018}
\setcopyright{acmcopyright}
\acmConference[SIGCOMM Posters and Demos '18]{ACM SIGCOMM 2018
Conference Posters and Demos}{August 20--25, 2018}{Budapest, Hungary}
\acmBooktitle{SIGCOMM Posters and Demos '18: ACM SIGCOMM 2018
Conference Posters and Demos, August 20--25, 2018, Budapest, Hungary}
\acmPrice{15.00}
\acmDOI{10.1145/3234200.3234201}
\acmISBN{978-1-4503-5915-3/18/08}




\maketitle

%\addtocounter{footnote}{1}
%\footnotetext[1]{Corresponding author.}

%\input{body}

%\input{introduction}

\vspace{-0.1cm}
\section{Background and Motivation}
\vspace{-0.1cm}

%\textbf{AR is gaining popularity;}
\subsection{Mobile IC apps are delay limited}
Emerging \emph{Immersive Computing (IC)} applications, such as virtual reality (VR) and augmented reality (AR), are changing the way human beings interact with mobile smart devices.
%Immersive functionalities are integrated on most popular mobile applications such as YouTube VR and Facebook Messenger, and are expected to increase in popularity as wearables and smart home devices continue to gain traction.
%\textbf{user-perceived latency is the key performance latency;}
%IC applications augment real scene via rendering high quality 3D annotations on top of user's view, or enable immersive experience by creating a virtual and interactive environment. To accomplish ultimate user experience, IC applications require very low user-perceived latency~\cite{lai2017furion}.
It is well-known that \emph{object recognition and rendering} are the key performance bottleneck in mobile IC systems~\cite{lai2017furion,zhang2018cars}. To speed up computation on mobile devices, the typical approach used in current IC applications is offloading computation-intensive tasks to the cloud~\cite{zhang2018cars,randeepdecision}, or leveraging local system optimizations to accelerate IC tasks~\cite{huynh2017deepmon,guo2018potluck}.
However, as user's QoE requirements increase over time, future mobile IC applications demand higher visual quality (\eg 4K or 8K resolution) and higher recognition accuracy/efficiency. %Running machine learning algorithms with high complexity or rendering high-quality immersive 3D effects efficiently involves significant computation/network overhead on hardware constrained mobile devices.
Exploring innovations working with existing offloading approaches and local optimizations to enhance the performance of future mobile IC applications is still an important but challenging problem facing the IC industry.

%due to the constrained hardware and network capabilities on mobile devices, above existing approaches trade off immersion quality for low latency. For example, on a mid-range Android smartphone, recognizing object from camera frames via inception DNN model takes several seconds per-frame~\cite{huynh2017deepmon}, and rendering one high-quality frame in a mobile VR system requires 100-200ms~\cite{lai2017furion}.

%\textbf{current solution: on-device or cloud-based AR, not enough;}
%\textbf{however current AR apps are not quick enough;}
%\textbf{look at the AR pipeline, recognition and rendering are bottleneck;}
%\textbf{optimizting AR performance is an important but challenging work;}

\vspace{-0.25cm}
\subsection{Computation redundancy in IC apps}
\vspace{-0.15cm}

To gain insight on how to further reduce the user-perceived latency in modern IC applications, we analyzed more than 30 popular mobile VR/AR applications collected from GooglePlay and AppStore to understand the user interactions and computation workload. We derived three IC-specific insights, indicating that \emph{IC tasks across different applications or users are often executed in similar or even redundant way}.

%what to do if current category is incorrect?

%\textbf{Recognition similarity:}

First, \emph{object recognition} is a typical computation-intensive approach used in mobile AR or other vision-based assistant applications. We observe that it usually processes similar inputs in different applications/users. For example, two safe-driving applications are likely to recognize the same stop sign from the different angle at the same crossroads.

%\textbf{Redundant 3D model rendering:}

Second, we observe that under certain circumstance, interactive VR/AR applications require rendering the same 3D model on various devices. If multiple users play in the same environment, the content in the view of different users is likely to be similar. For example, two Pokemon Go players require rendering the same 3D avatar when they are interacting through Pokemon application in the same place.

%\textbf{Redundant panoramas:}

In addition, current cloud-based VR applications leverage panoramic frames to create immersive experience~\cite{lai2017furion,boos2017flashback}. The server sends a panoramic frame to the client, and then the client crops the panorama to generate the final frame for display. Multiple users playing the same VR applications or watching the same VR video might use the same panorama.
% at the same time.

In summary, we argue that computation-intensive tasks of mobile IC applications can be similar or redundant, especially when applications/users are in the close location. The above insights suggest the opportunity to improve QoE of immersive computing by cooperatively sharing and utilizing intermediate IC results among different applications/users.

\vspace{-0.2cm}
\section{System Overview}
\vspace{-0.1cm}

\begin{figure}[tb]
\centering
\includegraphics[width=0.8\columnwidth]{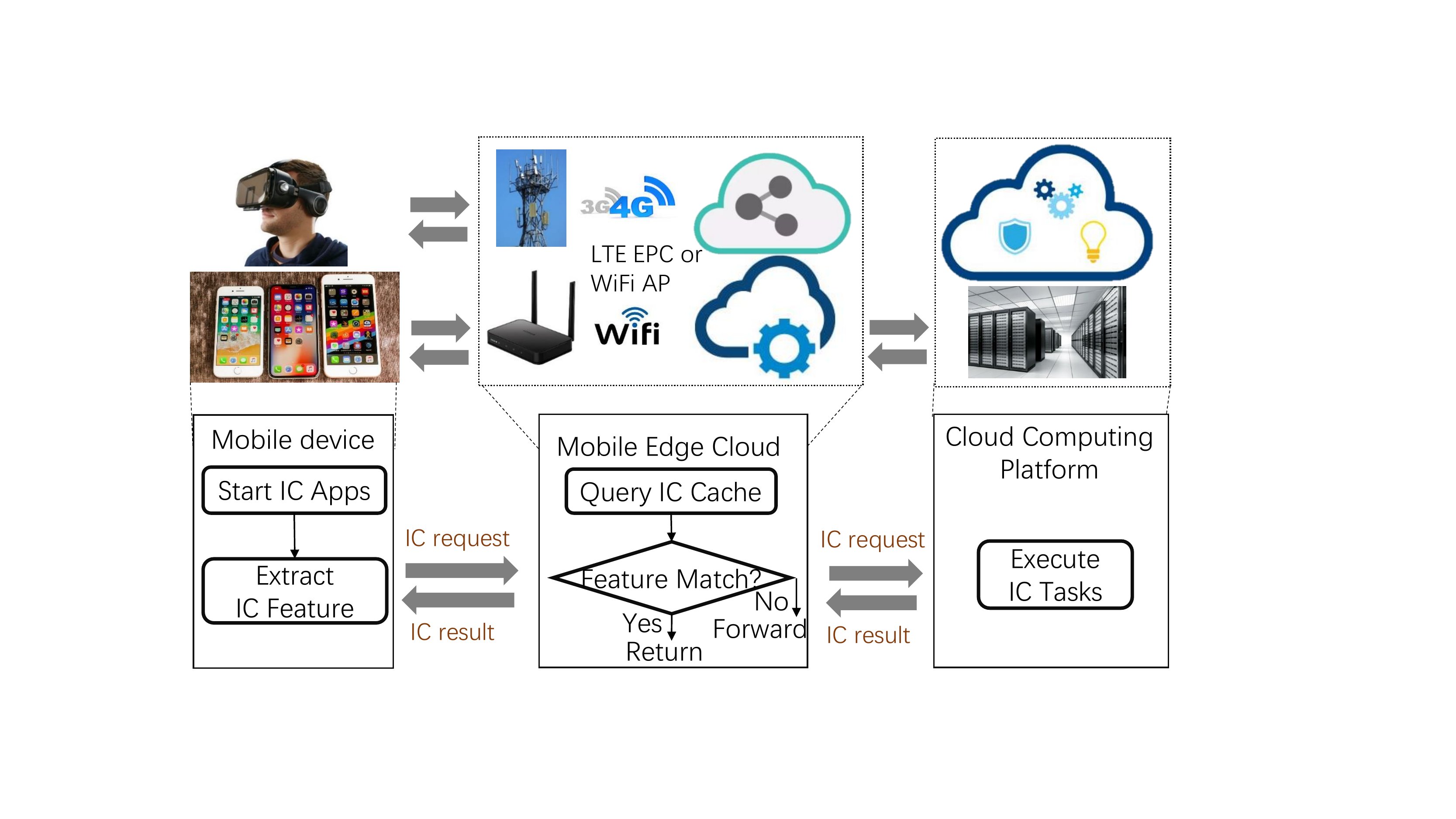}
\vspace{-0.1in}
\caption{\name architecture.}
\vspace{-0.3in}
\label{fig:architecture}
\end{figure}

We present \name, a cooperative framework for mobile immersive computing applications. To speed up computation-intensive IC tasks, \name leverages the insight that similar or redundant IC tasks among different applications/users can be cached and shared to improve the user-perceived quality of experience (QoE), especially the end-to-end latency.
%Several previous works have proposed to use deduplication/cache technologies to speed up computation intensive mobile applications~\cite{boos2017flashback,guo2018potluck,zhang2018cars}, but these works only focus on dedicated applications and do not take redundant tasks among different applications and users.
%At high-level, \name leverages the computation and storage capability on the mobile edge to identify and cache similar IC task in the same area. \name speeds up computation-intensive IC tasks by cooperatively sharing the result of similar or redundant IC tasks among different applications/users. Specifically, \name integrates the following key components on the mobile devices, edge, and cloud.
Figure~\ref{fig:architecture} shows the high-level architecture of \name.
%At high-level, \name is composed of
%(1) a client part that captures and sends IC request to the edge; %(1) a client lib that links to IC applications to transparently handle IC tasks;
%(2) a IC task cache on the edge to forward uncached request to the cloud, and cache intermediate computation result for cooperation; (3) a cloud engine to execute computation-intensive IC tasks (\eg recognition or rendering).
\if 0
\begin{table}[t]
  \centering
    \caption{\name's cache management policy. We define the key and match condition for each kind of IC tasks.}
  \label{tab:cache_management}
  \begin{tabular}{|p{2.3cm}|p{2.1cm}|p{2.7cm}|}
  \hline
  \textbf{IC Task} & \textbf{Feature (Key)} & \textbf{Match Condition} \\
  \hline
  \hline
  Object recognition & Convolutional feature & Distance is under threshold $\beta$ \\
  \hline
  3D Object rendering & Hash of 3D model & Hash value equals \\
  \hline
  VR video & Panoramic frame/video & Frame/video equals \\
  \hline
\end{tabular}
\end{table}
\fi
%The  workflow in \name is conceptually simple. 
Initially,
%the IC application calls IC APIs provided by \name's client lib to issue a IC request.
the client pre-processes the request to generate and send a \emph{feature descriptor} of user's input to the edge. On the edge, \name attempts to make a lookup with the feature descriptor (as the key) by matching the key to any results cached on the edge.
%a lookup attempt is made with the feature descriptor as the key and its corresponding task type by matching the input key to any existing key within a given similarity threshold.
If there is a hit, the cached result is returned to the client immediately. Otherwise, the edge forwards the request to the cloud and 
%then 
inserts the result
% returned from the cloud 
 to the edge cache.
%As shown in Table~\ref{tab:cache_management},

\name extracts dedicated property from each representative IC task as the feature descriptor.
%\name uses the pair of feature descriptor sets from the storage on the edge and the task request to identify a task match.
Specifically, for an object recognition task using DNN model, \name uses the feature vector generated from the input image as the feature descriptor. If the distance between the new feature descriptor and another one in the cache is under a certain threshold, \name determines that the computation result is already in the cache. For 3D object rendering and VR video streaming tasks, \name uses the hash value of the required 3D model or panoramic frames as the feature descriptor. Note that we did not cache object tracking results for AR applications because tracking is less computation-intensive as compared to recognition. Thus tracking is doable to be efficiently and accurately executed on mobile devices.

\vspace{-0.2cm}
\section{Preliminary results}
\vspace{-0.1cm}

\begin{figure}[t]%[htbp]
\centering
\vspace{-0.15cm}
\subfloat[Recognition latency reduction under different network conditions. $B_{M->E}$ and $B_{E->C}$ refer to the available bandwidth between mobile client and edge, edge and cloud, respectively.]{
\vspace{-0.15cm}
\label{fig:Recognition latency}
\begin{minipage}[t]{\columnwidth}%{0.79\columnwidth}
\centering
\includegraphics[width=0.74\columnwidth]{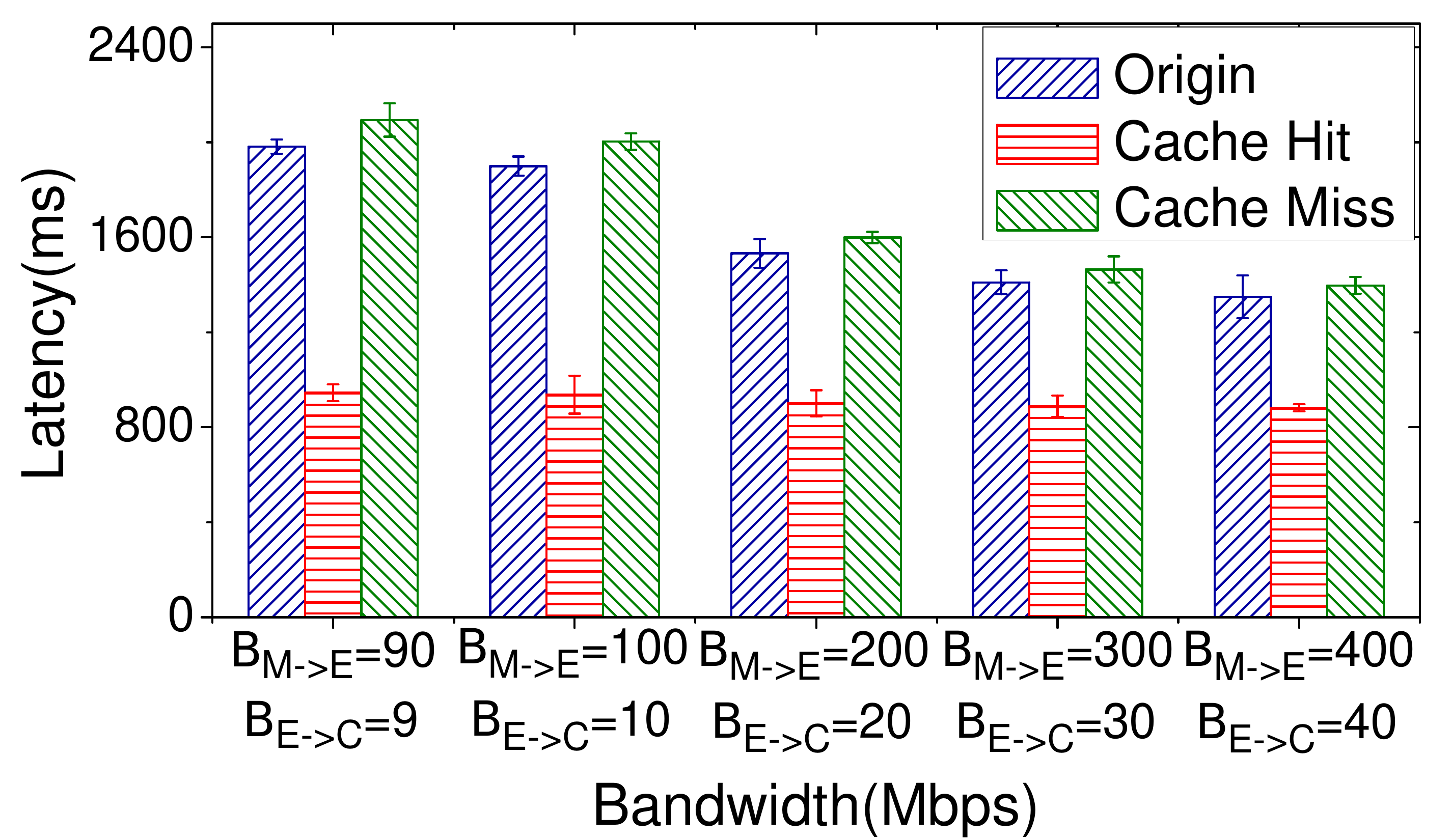}
\end{minipage}
}

\vspace{-0.15cm}
\subfloat[Load latency reduction in rendering tasks.]{
\vspace{-0.15cm}
\label{fig:Load latency}
\begin{minipage}[t]{\columnwidth}%{0.77\columnwidth}
\centering
\includegraphics[width=0.74\columnwidth]{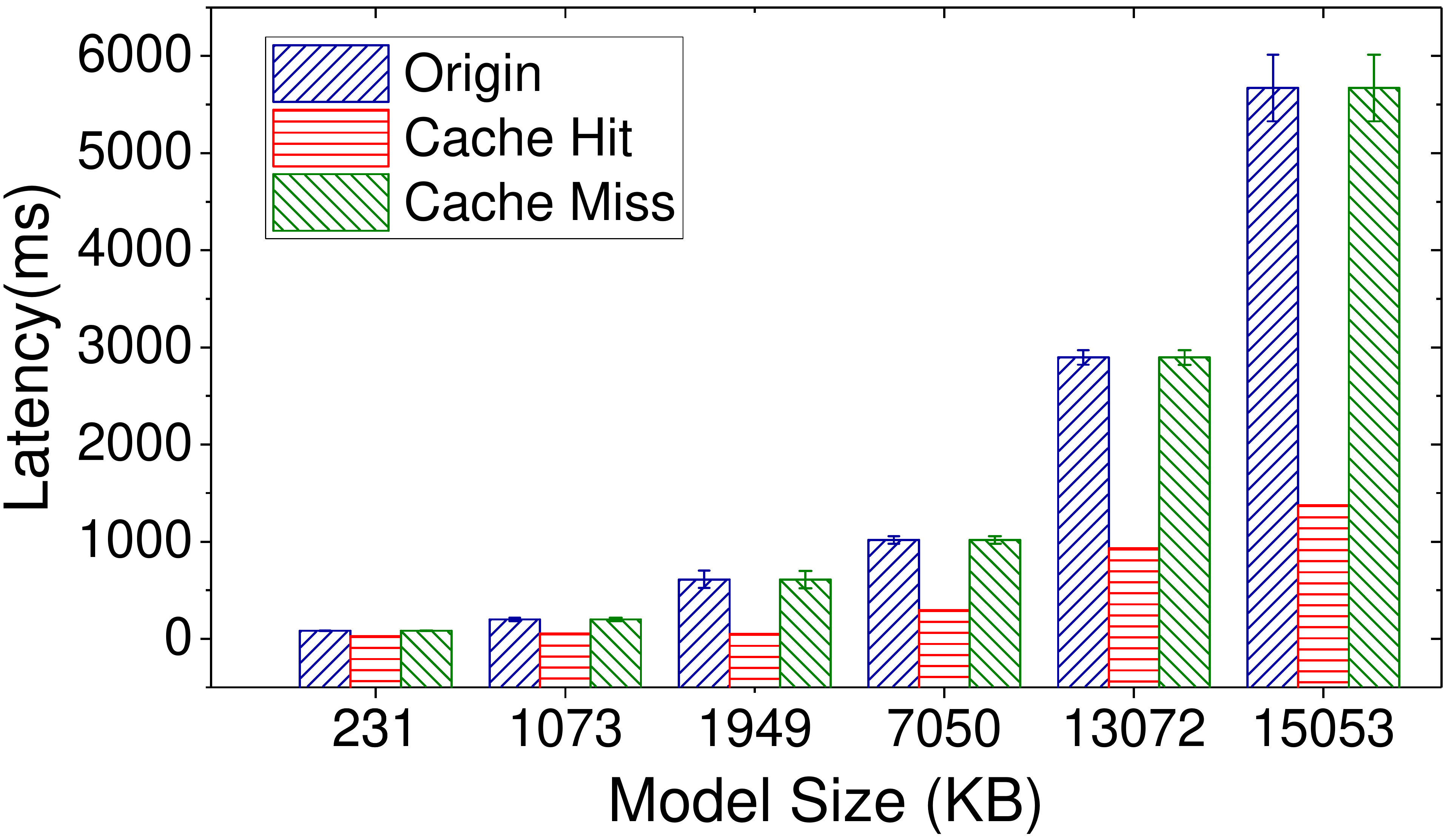}
\end{minipage}
}
\vspace{-0.3cm}
\caption{Latency reduction of computation-intensive tasks (object recognition and 3D model loading).}
\vspace{-0.7cm}
\label{fig:preliminary_results}
\end{figure}

To evaluate the QoE improvement by \name , we implement an AR application upon \name, which renders high-quality 3D annotations to label objects recognized in the camera view. The client part of \name is implemented on a Pixel smartphone running Android O, and the edge/cloud modules of \name are implemented on two Linux machines respectively. The client connects to the edge via 802.11ac WiFi which supports up to 400Mbps available throughput in our experiment. We use \texttt{tc}
%(a Linux command tool)
to tune the network condition to simulate real wireless/mobile network.
%We evaluate the recognition latency and rendering initiation time of \name, as shown in Figure~\ref{?}.
In our experiment, the client sends recognition and rendering requests to the edge. The edge returns results immediately if the result is found in the cache.
 Otherwise the edge forwards the request to the server.
 We use an origin version which offloads complete IC tasks to the cloud without cache as the baseline. Figure~\ref{fig:Recognition latency} plots the reduction of recognition latency by \name. Current \name implementation recognizes objects via a DNN model. By identifying and caching similar computation results of the DNN model, \name can reduce up to 52.28\% recognition latency under different network conditions. Figure~\ref{fig:Load latency} plots the latency reduction in rendering tasks. To execute a rendering task, the renderer has to load the 3D model into memory first and draw objects on the display. By caching the loaded data in rendering tasks on the edge, \name reduces the load latency by up to 75.86\% for 3D models differed in size.

\vspace{-0.1cm}
\section{Ongoing and future work}
\vspace{-0.1cm}
Our end goal is to improve the QoE of mobile IC applications by cooperatively utilizing the similar/redundant IC workload among different applications/users. Since the current \name can only identify coarse-grained IC tasks with simple cache management policy, we are exploring the improvement that can efficiently and accurately identify reusable IC workload in fine-grained (\eg the result of a specific DNN layer). In addition, we will also study on the security/privacy protection issues in the cooperative system.

%1) dynamic threshold; 2) DNN partition; 3) security issue.

%\input{conclusion} 

\vspace{-0.1cm}
\bibliographystyle{ACM-Reference-Format}%{abbrv}%
\bibliography{reference}

\end{document}